\documentclass[12pt]{iopart}

\usepackage{iopams}
\usepackage{graphicx}  
\begin{document}

\title{Collision of two Hopfions}

\author{M Array\'as and J L Trueba}

\address{\'Area de Electrogmanetismo, Universidad Rey Juan Carlos, Tulip\'
an s/n, 28933 M\'ostoles, Madrid, Spain}
\ead{manuel.arrayas@urjc.es, joseluis.trueba@urjc.es}
\vspace{10pt}
\begin{indented}
\item[]October 10th 2016
\end{indented}

\begin{abstract}
We study the collision of two hopfions or Hopf-Ra\~nada electromagnetic fields. The superposition of two of such fields, travelling in opposite directions, yields different topology for the electric and magnetic field lines. Controlling the angular momentum of such fields, we can control the topology of the flow associated to the field lines, and the energy distribution. The concept of electromagnetic helicity and the exchange between its magnetic and electric components are used to explain the different behaviours observed when the angular momentum is reversed.
\end{abstract}

%
\vspace{2pc}

\submitto{\jpa}
%
%
%

\section{Introduction}
\begin{figure}[h]
    \begin{center}
       \includegraphics[width=0.43\linewidth]{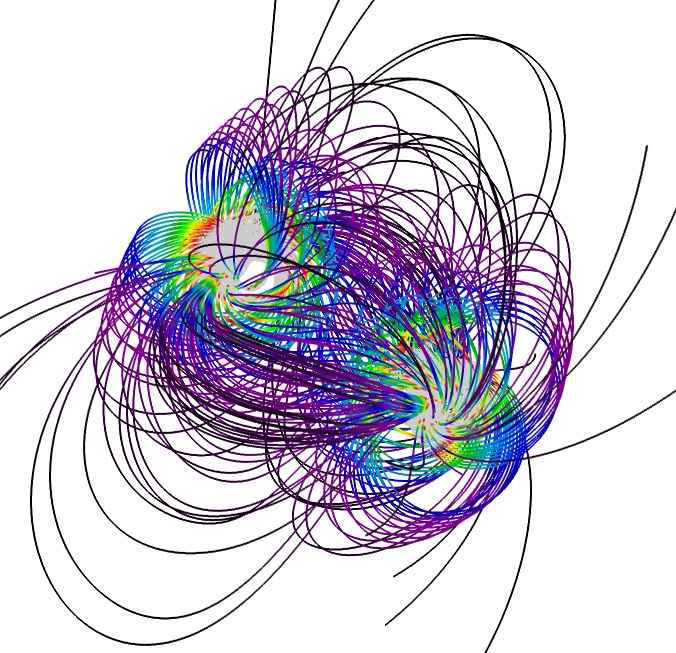}
       \includegraphics[width=0.43\linewidth]{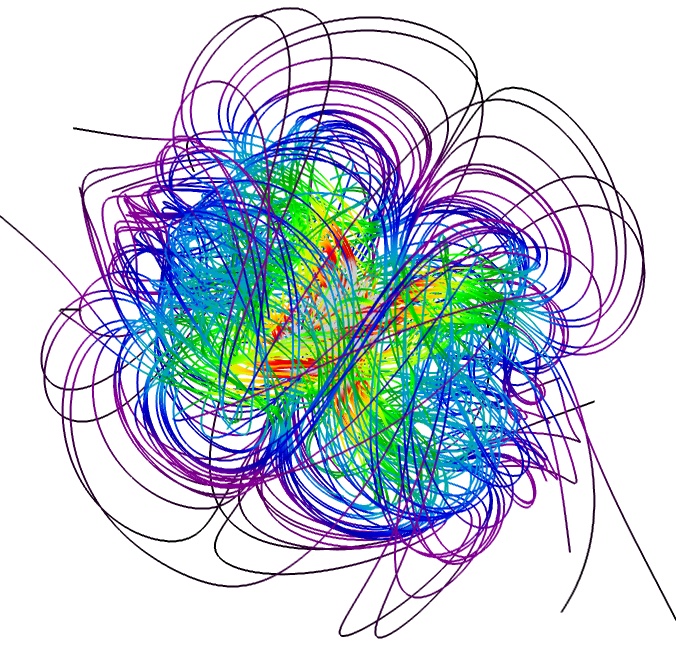}\\
       \includegraphics[width=0.43\linewidth]{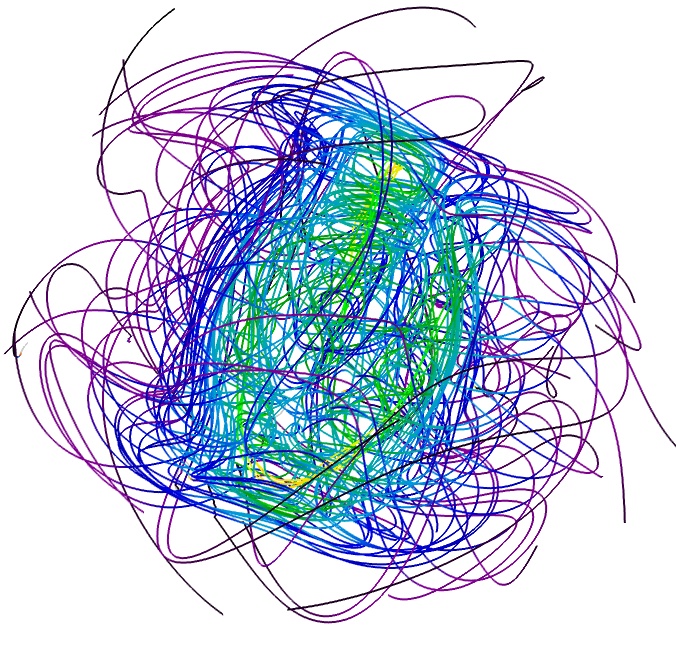}
       \includegraphics[width=0.43\linewidth]{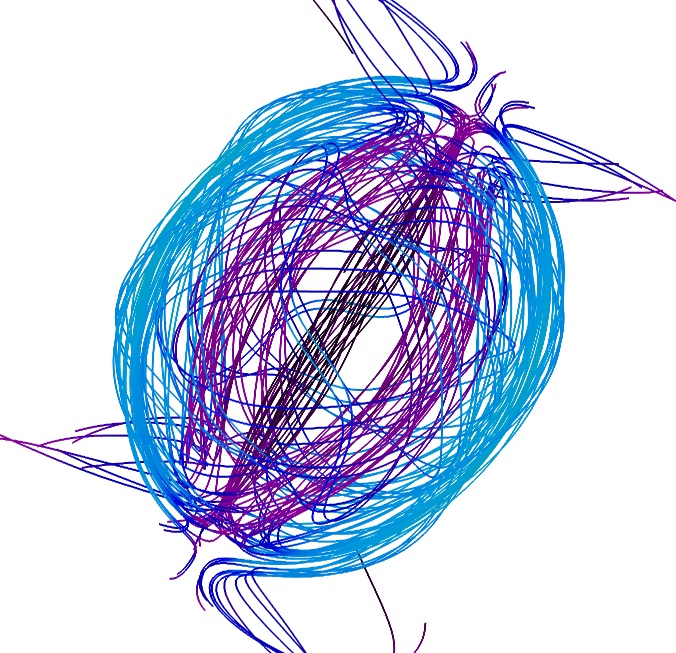}
       \caption{Magnetic field lines for the collision of hopfions with null total angular momentum (${\bf J}=0$), from left to right and top to bottom, at times $T=0, 2, 4, 6$.}
       \label{lines}
    \end{center}
\end{figure}

\begin{figure}[h]
    \begin{center}
       \includegraphics[width=0.43\linewidth]{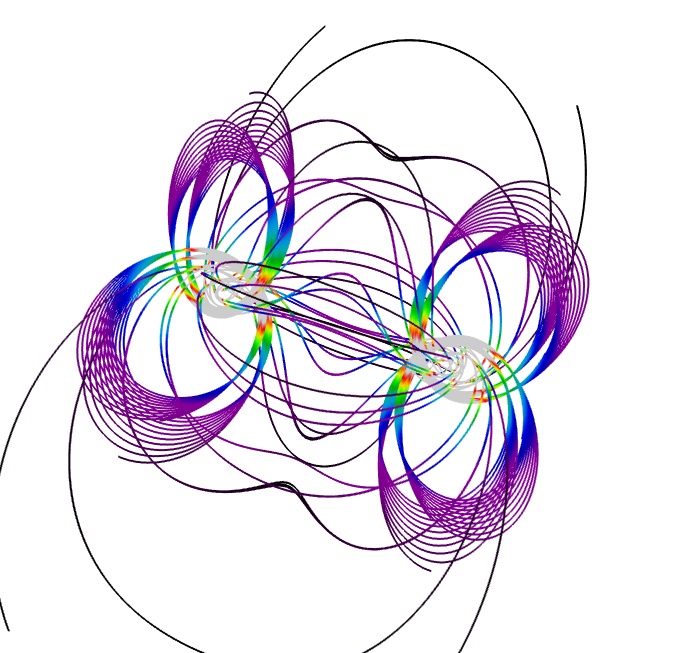}
       \includegraphics[width=0.43\linewidth]{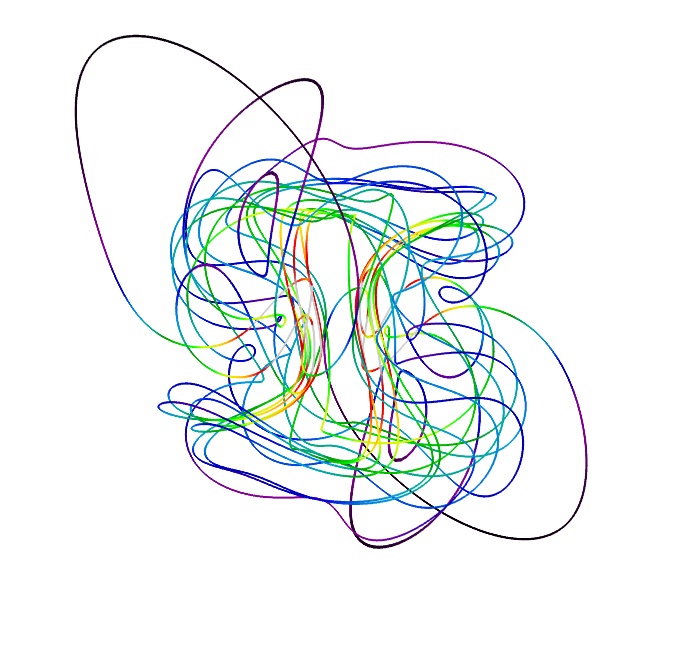}\\
       \includegraphics[width=0.43\linewidth]{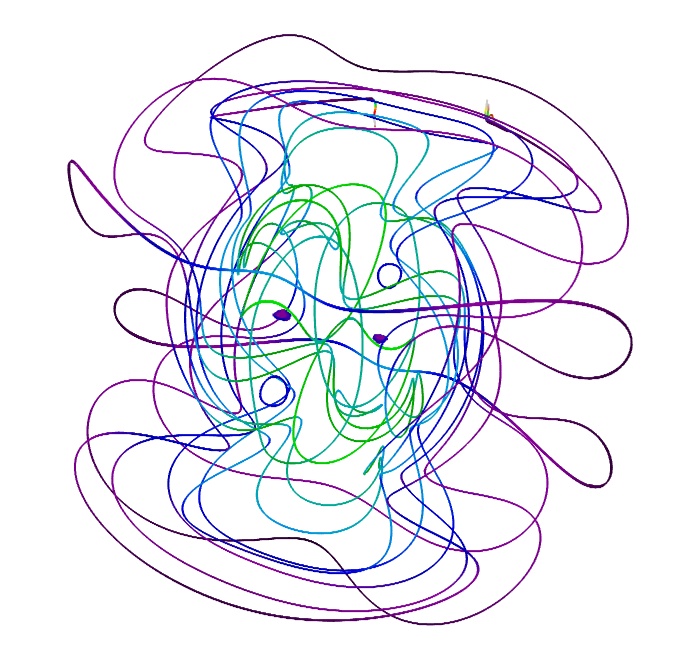}
       \includegraphics[width=0.43\linewidth]{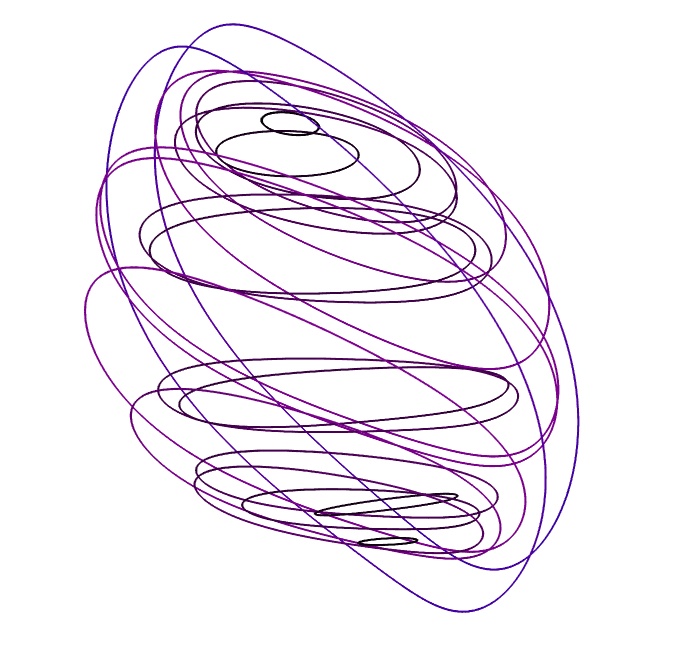}
\caption{Magnetic field lines for the collision of hopfions with a net total angular momentum (${\bf J}\neq 0$), from left to right and top to bottom, at times $T=0, 2, 4, 6$.}
        \label{linesn}
    \end{center}
\end{figure}

A hopfion or Hopf-Ra\~nada field is an example of electromagnetic knot. It is a solution of Maxwell's equations in vacuum \cite{Ran89,Ran90} where the electric and magnetic field lines form closed loops each of them linked to another. Such fields have also appeared as solitons in MHD \cite{Kamchatnov} and generalised  later \cite{Bouw2014,Bouw2015}. The hopfion is deeply connected to non-null twistors \cite{Dalh2012,Bouw2015b} and its properties have been studied \cite{Irv08,Arr10} extensively. In this work we consider the collision or interference of two hopfions and show how we can change the topology of the flow associated to the field lines by changing the angular momentum. We take two hopfions, initially focused at two different points in the space and travelling towards each other along the line joining the focused points. In one case we make them interfere with opposite angular momentum, and in the other case they do carrying the same angular momentum. Thus in the first case the total angular momentum of the electromagnetic field is zero (${\bf J}=0$), while in the second is non-null (${\bf J}\neq 0$). In \Fref{lines} for the first case some magnetic lines are plotted for a total angular momentum equals zero at different times and we can see a very strong entanglement of the field lines. When the total angular momentum is different to zero, the flow changes and the entanglement of the field lines is modified as seen in \Fref{linesn}. As a result, the energy flux distribution is also modified (see \Fref{ison} and \Fref{iso}). We will provide an explanation of the effects of the angular momentum in the collision and prove that the concept of the helicity and the exchange mechanism \cite{Arr12} is useful to understand the topological behaviour of the field lines. The results presented here may be of use when dealing with the interaction of other solitons-like structures appearing in superfluid Helium \cite{Volovik1977}, Bose-Einstein condensates \cite{Kawaguchi2008} and ferromagnetic materials \cite{Dzyloshinskii}.

\section{The colliding hopfion fields}

The hopfions can be written in terms of two complex scalar maps \cite{Ran97},
\begin{equation}
  \phi={z_0}/{z_1}, \,\,\,\,\,\,\,\,\, \theta = {w_0}/{w_1},
\label{eq:pp}
\end{equation}
being
\begin{eqnarray}
z_0&=& \alpha + i \beta,\,\,\,  z_1= \gamma + i \delta,\nonumber\\
w_0&=& \beta+ i \gamma,\,\,\, w_1= \alpha + i \delta,
\label{knot23}
\end{eqnarray}
and 
\begin{eqnarray}
\alpha &=& \frac{A \, X - T \, Z}{A^2+T^2} , \,\,\,\,\,\,\, \beta = \frac{A \, Y + T \, (A-1)}{A^2+T^2}\nonumber\\
\gamma&=& \frac{A \, Z + T \, X}{A^2+T^2}, \,\,\,\, \,\,\, \delta = \frac{A \, (A-1) - T \, Y}{A^2+T^2}.
\label{coordS3}
\end{eqnarray}
The coordinates $(T,X,Y,Z)=(ct,x,y,z)/L_0$ are dimensionless Minkowskian coordinates and $A = (R^{2}-T^{2}+1)/2$, with $R^2=X^2+Y^2+Z^2$. The length $L_0$ represents the mean quadratic radius of the energy density distribution at $T=0$ of the hopfion field \cite{Arr10}. 

The quantities $(\alpha,\beta,\gamma,\delta)$ defined in \eref{coordS3} satisfy $\alpha^2+\beta^2+\gamma^2+\delta^2=1$ and represent time-dependent real coordinates of the three-dimensional sphere $S^3$. Using the stereographic projection, the complex scalar field $\phi$ in \eref{eq:pp} can be written as real coordinates $(n_1,n_2,n_3)$ in the two-dimensional sphere $S^2$, where $n_1^2+n_2^2+n_3^2=1$. Thus $\phi = (n_1+in_2)/(1-n_3)$, with $i$ the imaginary unit, and the same can be done for $\theta$. Therefore the expressions for the hopfion given by \eref{eq:pp} constitute time-dependent maps $S^3\to S^2$. The topology of the level curves of a map $S^3\to S^2$ is characterised by a Hopf index \cite{Hop31}. The Hopf index coincides with the Gauss linking number \cite{Ricca2011} of any two different level curves of the map. For the hopfion given by \eref{eq:pp}-\eref{coordS3} the Hopf index of $\phi$ and $\theta$ is equal to 1 at any time. 

The magnetic and electric fields associated to such maps can be obtained using the Ra\~nada construction \cite{Ran89,Ran90}, 
\begin{eqnarray}
{\bf B}({\bf r},t) &=& \frac{\sqrt{a}}{2\pi i} \, \frac{\nabla \phi \times \nabla \bar{\phi}}{(1+ \bar{\phi} \phi )^{2}}, \label{ranknot1intro} \\
{\bf E}({\bf r},t) &=& \frac{c\sqrt{a}}{2\pi i} \, \frac{\nabla \bar{\theta} \times \nabla \theta}{(1+ \bar{\theta} \theta)^{2}},\label{ranknot2intro}
\end{eqnarray}
where $\bar{\phi}$ and $\bar{\theta}$ denote the complex conjugates, $a$ is a constant introduced so the fields have correct dimensions, and $c$ the speed of light in vacuum. By construction, the field lines coincide with the level curves of the maps $\phi$ and $\theta$.

The Hopf-Ra\~nada fields transport linear and angular momentum along the $y$-axis. The linear and angular values corresponding to the maps given by \eref{eq:pp} are ${\bf p}_1=(a/c\mu_0L_0){\bf u}_y$ and ${\bf J}_1 =(a/c\mu_0){\bf u}_y$ respectively, being $\mu_0$ the vacuum magnetic permeability.  We will use the subscripts $(++)$ to denote applications with positive ${\bf p}$ and ${\bf J}$ along the $y$-axis. Then, the applications given by \eref{eq:pp} at $T=0$ with positive linear and angular momentum along the $y$-axis will be denoted from now on as $\phi_{++},\theta_{++}$.

In order to study the collision of two hopfions, we need to reverse the linear momentum of one of them. We can do that by a reflection transform in $S^3$ at $T=0$. To study the effect of changing the angular momentum in the collision, a further reflection can be applied at $T=0$. Thus we obtain,
\begin{eqnarray}
  \phi_{-+}=\frac{-{\bar z}_0}{z_1}, \,\,\,\,\,\,\,\,\, \theta_{-+} = \frac{w_0}{-{\bar w}_1},\label{eq:mp}\\
  \phi_{--}=\frac{-{\bar z}_0}{{\bar z}_1}, \,\,\,\,\,\,\,\,\, \theta_{--} = \frac{{w}_0}{-{ w}_1}.\label{eq:mm}
\end{eqnarray} 
There is another transform we will make use which is a translation $\mathcal{T}_{y_0}f({\bf r})=f({\bf r}+y_0{\bf u}_y)$. With that translation we can control the focus of the hopfions at the initial time.

\section{Collisions with different angular momentum}

\begin{figure}
    \begin{center}
       \includegraphics[width=0.43\linewidth]{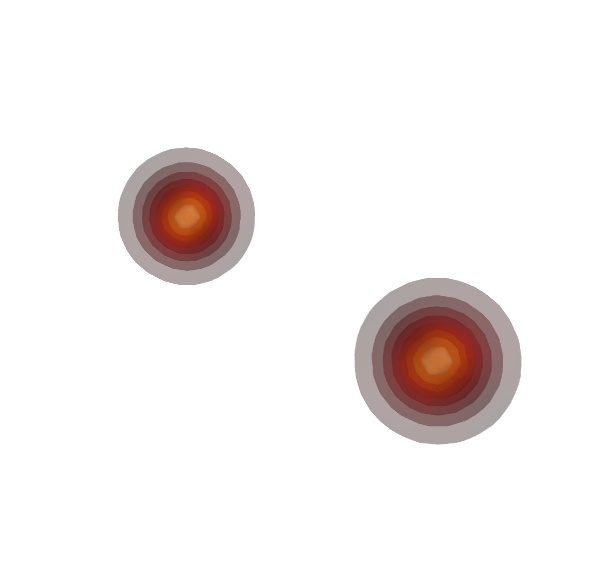}
       \includegraphics[width=0.43\linewidth]{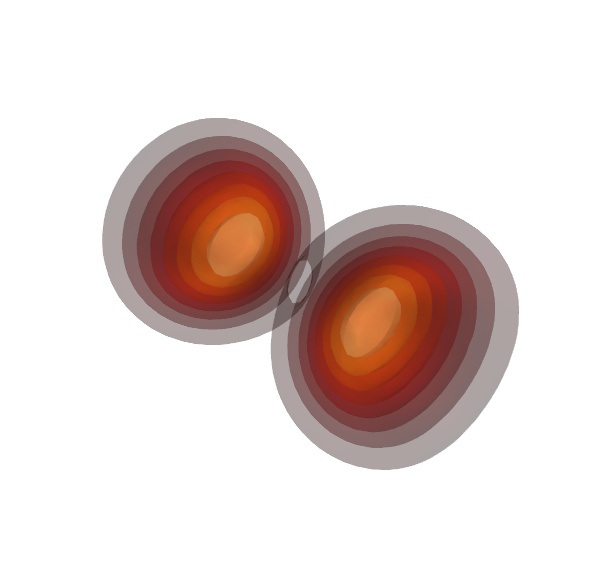}\\
       \includegraphics[width=0.43\linewidth]{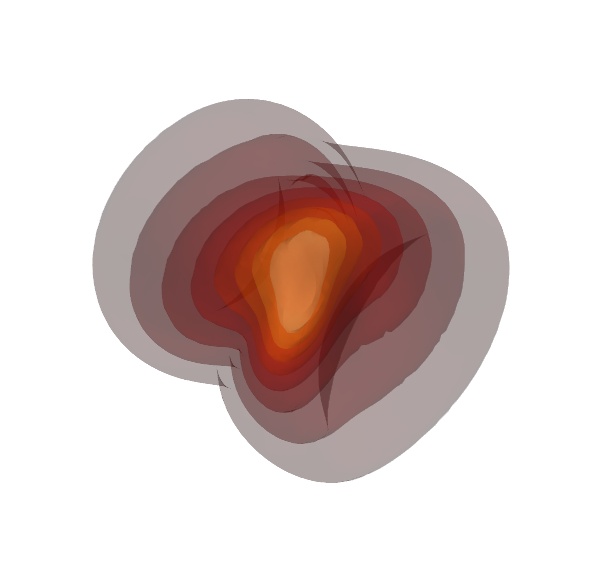}
       \includegraphics[width=0.43\linewidth]{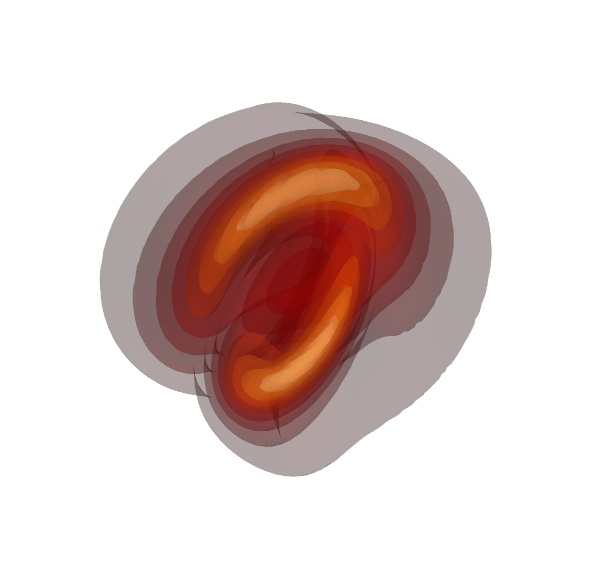}
       \caption{The collision of two hopfions given by $\mathcal{T}_{-2}(\phi_{++},\theta_{++})$, and $\mathcal{T}_{+2}(\phi_{--},\theta_{--})$ respectively. Isosurfaces of constant energy, from left to right and top to bottom, at times $T=0, 1, 2, 3$. Initially the hopfions are focused at  $(X_0,Y_0,Z_0)=(0, \pm 2, 0)$ (left-top corner). They travel in opposite directions along the $y$-axis.}
       \label{ison}
    \end{center}
\end{figure}

\begin{figure}
    \begin{center}
       \includegraphics[width=0.43\linewidth]{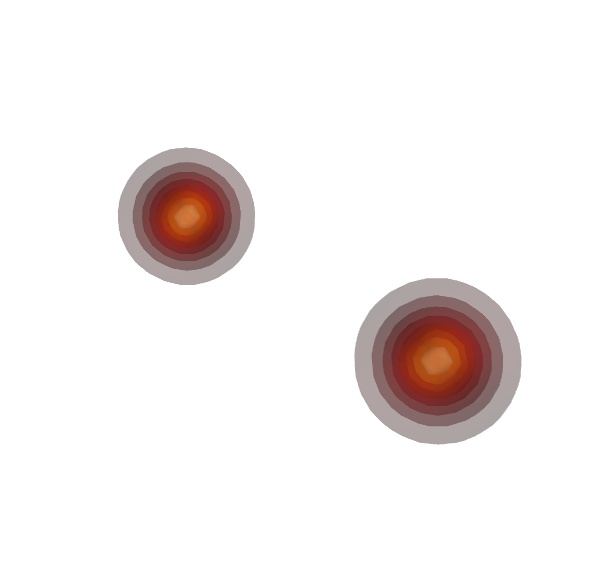}
       \includegraphics[width=0.43\linewidth]{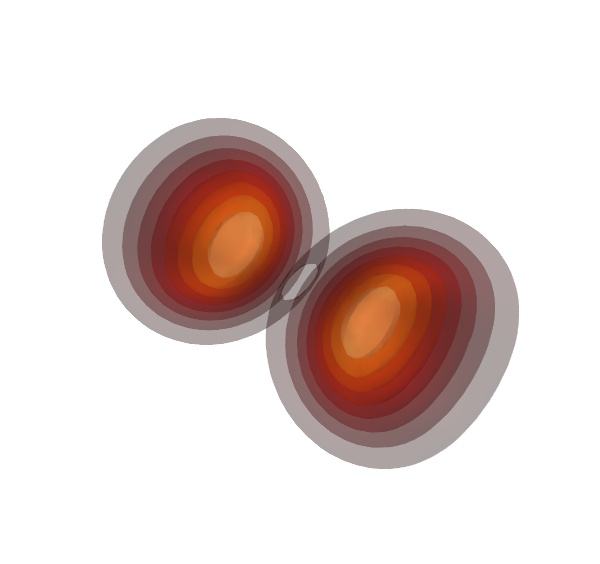}\\
       \includegraphics[width=0.43\linewidth]{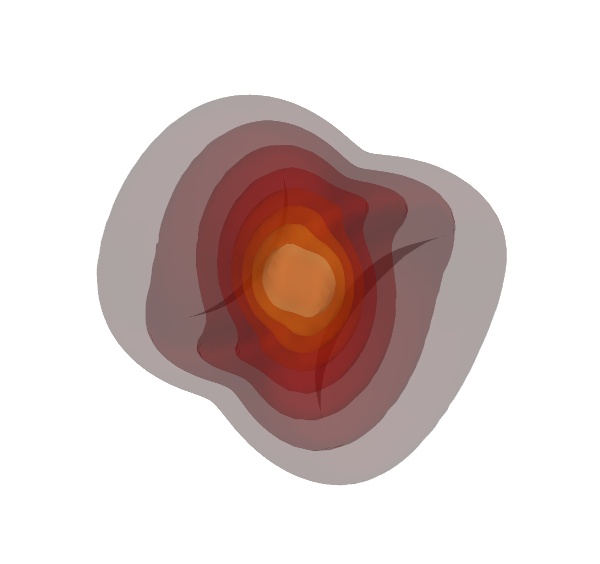}
       \includegraphics[width=0.43\linewidth]{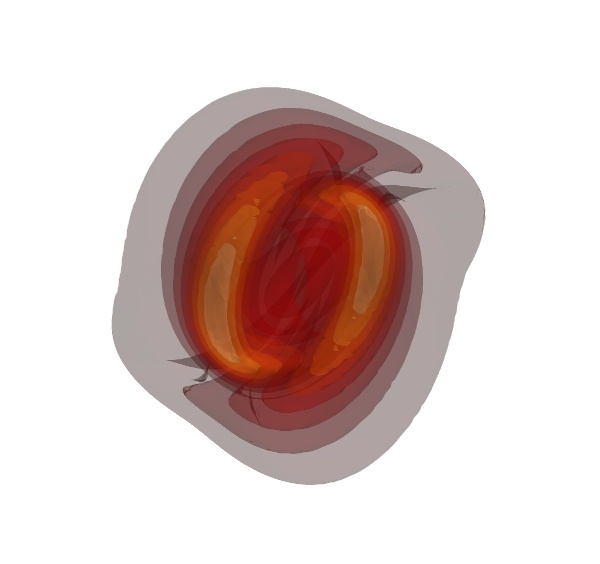}
       \caption{The collision of two hopfions given by $\mathcal{T}_{-2}(\phi_{++},\theta_{++})$, and $\mathcal{T}_{+2}(\phi_{-+},\theta_{-+})$ respectively. Isosurfaces of constant energy, from left to right and top to bottom, at times $T=0, 1, 2, 3$. Initially the hopfions are focused at  $(X_0,Y_0,Z_0)=(0, \pm 2, 0)$ (left-top corner). They travel in opposite directions along the $y$-axis.}
       \label{iso}
    \end{center}
\end{figure}

 We consider the collision of two hopfions initially focused at $Y=\pm 2$ respectively. As considering a collision, they will have opposite linear momentum along the $y$-axis, so the total linear momentum will be zero, ${\bf p} =0$. In the first case, the hopfions are given by $\mathcal{T}_{-2}(\phi_{++},\theta_{++})$ and $\mathcal{T}_{+2}(\phi_{--},\theta_{--})$, thus the total field is characterised by a total angular momentum ${\bf J}= 0$. In the second case, the hopfions are given by $\mathcal{T}_{-2}(\phi_{++},\theta_{++})$ and $\mathcal{T}_{+2}(\phi_{-+},\theta_{-+})$, and the total field has ${\bf J}= 2{\bf J}_1= (2a/c\mu_0){\bf u}_y$. In \Fref{ison} it is plotted the first case. The picture show isosurfaces of constant energy, from left to right and top to bottom, at times $T=0, 1, 2, 3$. At $T=0$ the hopfions are focused at  $(X_0,Y_0,Z_0)=(0, \pm 2, 0)$ as can be seen at the right-top corner of the figure. They travel in opposite directions along the $y$-axis with opposite angular momentum. The second case is shown in \Fref{iso}. The difference is that in this case they travel along the $y$-axis with opposite linear momentum as before but now they have the same angular momentum pointing in the same direction. It is observed that the interference pattern changes and so it does the distribution of the energy in the space as the isosurfaces of constant energy show when comparing both cases.  
  

The differences of the isosurfaces plotted in \Fref{ison} and \Fref{iso} are appreciated when the energy centres of the hopfions approach to each other and the interference is bigger. The shapes observed in each case are a consequence of the individual angular momentum of the colliding hopfions. In \Fref{ison} the rotational direction of the Poynting vector of one hopfion is counterclockwise, while for the other hopfion is clockwise, so when they encounter in the collisional region, the isosurfaces appear twisted. This is the case of a total angular momentum equal zero as ${\bf J}={\bf J}_1-{\bf J}_1=0$. In \Fref{iso} the rotational direction of the Poynting vector is the same for the individual hopfions, so when they encounter in the collisional region, the isosurfaces appear without twist. This is the case of a total angular momentum equal zero as ${\bf J}=2{\bf J}_1$. In the energy isosurface plots, the superposition of the two hopfions, can be visualised as the collision of two individual particles. The individual character can be appreciated at the earlier time, when the energy density is concentrated at the focused points $Y=\pm 2$, travelling to each other. The individual character is lost when the centres are close.

\section{Interchange of helicity and flow topology}

A measure of the degree of entanglement of the field lines is provided by the helicity of the field \cite{Calugareanu,Moffatt,Berger}. The electromagnetic helicity $h$ \cite{True96,Arr12} is defined for electromagnetic fields in vacuum as the sum of the magnetic and electric helicities, 
\begin{equation}
h = h_{m} + h_{e} .
\label{eq:h}
\end{equation}
The magnetic helicity, that can be written in SI units as
\begin{equation}
h_{m} = \frac{1}{2c\mu_0}\int d^3r\, {\bf A}\cdot{\bf B} ,
\label{maghel}
\end{equation}
where ${\bf B}=\nabla \times {\bf A}$, is proportional to the mean Gauss linking number of the magnetic lines. Similarly, the electric helicity,
\begin{equation}
h_{e} = \frac{\varepsilon_0}{2c}\int d^3r\, {\bf C}\cdot{\bf E},
\label{elhel}
\end{equation}
where ${\bf E}=\nabla \times {\bf C}$, is proportional to the mean Gauss linking number of the electric lines and can only be defined for vacuum fields. Note that, by definitions (\ref{maghel}) and (\ref{elhel}), both helicities have dimensions of angular momentum as the helicity in particle physics.

One of the intriguing properties of the hopfion defined in equations (\ref{eq:pp}-\ref{ranknot2intro}) is the fact that, for every time, the Gauss linking number of any pair of magnetic lines is equal to 1, and the Gauss linking number of any pair of electric lines is equal to 1 \cite{Ran01}. Accordingly, the hopfion satisfies for every time the relations
\begin{equation}
h_{m} = h_{e} = \frac{a}{2 c \mu_{0}} ,
\label{hopfhel1}
\end{equation}
and
\begin{equation}
h = h_{m} + h_{e} = \frac{a}{c \mu_{0}} .
\label{hopfhel2}
\end{equation}
In these equations, the constant $a/ 2 c \mu_{0}$ can be understood as a unit of helicity.

In particle physics, the helicity of a photon can be seen as the projection of its angular momentum along the axis given by its linear momentum,
\begin{equation}
h_{photon} = \frac{{\bf J} \cdot {\bf p}}{p}, 
\label{photon}
\end{equation}
where $p$ is the modulus of the linear momentum ${\bf p}$ and ${\bf J}$ is the angular momentum. Therefore by reversing the linear momentum of the photon, or by reversing its angular momentum, the helicity of the photon (\ref{photon}) changes its sign. The hopfions studied in the present work behaves in a similar way. We have considered the collision of two Hopfions with opposed linear momenta. In a first case, we have also reversed the angular momentum of the second hopfion. For this first case each colliding hopfions satisfy
\begin{equation}
h_{m} (+,+) = h_{e} (+,+) = h_{m} (-,-) = h_{e} (-,-) = \frac{a}{2 c \mu_{0}} ,
\label{hopghel3}
\end{equation}
 Then the helicity of each hopfions is
\begin{equation}
h (+,+) = h (-,-) = \frac{a}{c \mu_{0}} ,
\label{hopfhel4}
\end{equation}
which shows a similar behaviour as the particle helicity (\ref{photon}). In a second case, we have taken two hopfions colliding with opposed linear momentum but the same angular momentum. These hopfions have
\begin{equation}
h_{m} (+,+) = h_{e} (+,+) = \frac{a}{2 c \mu_{0}} , \; h_{m} (-,+) = h_{e} (-,+) = - \frac{a}{2 c \mu_{0}} ,
\label{hopghel5}
\end{equation}
and 
\begin{equation}
h (+,+) = \frac{a}{c \mu_{0}} , \; h (-,+) = - \frac{a}{c \mu_{0}} ,
\label{hopfhel6}
\end{equation}
again similar to (\ref{photon}).

The superposition of two electromagnetic fields with opposed linear momentum in the collision shows in principle interference phenomena. In the collision of the two hopfions, an interesting fact is that the electromagnetic helicity (as well as the linear and angular momentum) behaves as if there is no interference, and we have
\begin{equation}
h = h (p_1,J_1) + h (p_2,J_2) .
\label{hopfhel7}
\end{equation}
Thus in the first case, with opposed linear and angular momenta, we have a total field with no linear and angular momentum, but with a net electromagnetic helicity given by $h = 2a/(c\mu_0)$. On the other hand, for the second case, with only opposed linear momentum, the electromagnetic helicity turns out to be null $h=0$. This non-interference behaviour is not so for the magnetic and the electric helicities, as we will see below. 

Since the electromagnetic helicity is related to the linking number of field lines, we could then expect that the field lines in the first case (${\bf J}=0$ and $h \neq 0$), are going to be entangled, or arrange in a knotted mess, while in the second case (${\bf J}\neq 0$ and $h = 0$) they are not. In order to check that hypothesis, we have plotted in \Fref{lines} and \Fref{linesn} few magnetic field lines at different times for both cases. 

It is observed that in \Fref{lines}, representing the case $h \neq 0$, the field lines are tangled forming a kind of ball when time evolves. In the figures, entanglement is easy to see but the actual linkage of magnetic lines, or the fact that it may or not depend on time, needs more explanation. The magnetic helicity, being related to the mean linking number of the magnetic lines, can be used to make some progress on such matters. The electromagnetic helicity of every electromagnetic field in vacuum is constant in time \cite{True96}. However, the magnetic and electric helicities may change with time, only their sum being constant, in such way that it is possible an exchange of helicity between the magnetic and the electric fields \cite{Arr12} in the following way,
\begin{equation}
\frac{d h_{m}}{dt } = - \frac{dh_{e}}{dt} = - \frac{1}{c \mu_{0}} \int d^3 r \, {\bf E} \cdot {\bf B} ,
\label{hopfhel8}
\end{equation}
in which it is assumed that fields and related potentials are null at infinity. The individual hopfions have constant magnetic and electric helicities because the integral (\ref{hopfhel8}) is zero for both of them. However, in the collision, interference occurs and the total electric and magnetic fields can have a non-zero value of that integral. In the case of a collision with total angular momentum ${\bf J}=0$ and total electromagnetic helicity $h \neq 0$, the initial values of the magnetic and electric helicities are
\begin{eqnarray}
h_{m} (T=0) &=& h_{m} (+,+) + h_{m} (-,-) = \frac{a}{c \mu_{0}} , \label{hopfhel9} \\ 
h_{e} (T=0) &=& h_{e} (+,+) + h_{e} (-,-) = \frac{a}{c \mu_{0}} .
\label{hopfhel10}
\end{eqnarray}
Thus at $T = 0$, there is a non-null mean value of the linking number of the magnetic lines, and the same value of the linking number of electric lines. The magnetic case at $T=0$ is observed in the first plot of \Fref{lines}. We have computed numerically the integral (\ref{hopfhel8}) at $T = 0$, obtaining 
\begin{equation}
\left. \frac{d h_{m}}{dt} \right|_{T=0} = \left. - \frac{dh_{e}}{dt} \right|_{T=0} = - \frac{1}{c \mu_{0}} \, 0.0064 .
\label{hopfhel11}
\end{equation}
Accordingly, the magnetic helicity decreases immediately after $t = 0$ and the electric helicity increases in the same quantity. The mean linking number of the magnetic lines has to decrease very close to $t = 0$. A similar computation for $T=2$ gives the result
\begin{equation}
\left. \frac{d h_{m}}{dt} \right|_{T=2} = \left. - \frac{dh_{e}}{dt} \right|_{T=2} = \frac{1}{c \mu_{0}} \, 0.7336 ,
\label{hopfhel12}
\end{equation}
so the magnetic helicity now increases and the electric helicity decreases. Then, the mean value of the linking number in \Fref{lines} depends on time.

Consider now the second case, with ${\bf J} \neq 0$ and $h = 0$, represented in \Fref{linesn}. In this case, the initial values of the total magnetic and electric helicities are
\begin{eqnarray}
h_{m} (t=0) &=& h_{m} (+,+) + h_{m} (-,+) = 0 , \label{hopfhel13} \\ 
h_{e} (t=0) &=& h_{e} (+,+) + h_{e} (-,+) = 0 .
\label{hopfhel14}
\end{eqnarray}
The mean linking number of the magnetic lines, and also of the electric lines at $t = 0$ is null. However, some of the magnetic lines at $t=0$ shown in the first panel of \Fref{linesn} seem to be linked. To understand this behaviour, we may look at the similar linking pattern of the lines at the left and the right of the first panel of \Fref{linesn}. In some of these lines, we can see that the sign of the linking number at the left is opposed to the sign of the linking number at the right. The mean value of the linking number is then equal to zero even if there is linkage of pairs of lines. Numerical integration of the integral (\ref{hopfhel8}) shows that, in the case of null total electromagnetic helicity, at any time 
\begin{equation}
\left. \frac{d h_{m}}{dt} \right| = \left. - \frac{dh_{e}}{dt} \right| = 0 .
\label{hopfhel15}
\end{equation}
Consequently, the total magnetic and electric helicities are constant during the collision and, using the results (\ref{hopfhel13}-\ref{hopfhel14}) they are null at every time. There is no exchange of helicities in this case and all the examples shown in \Fref{linesn} have to correspond to a mean linking number equal to zero.

\section{Conclusions}

In conclusion, in this paper we have considered the collision of two hopfions, initially focused at two different points in the space and travelling towards each other along the line which joins the initial focused points. We have observed that topology of the flow associated to the field lines changes whether the hopfions have opposite or equal angular momentum. The energy distribution in the space is also modified. 

The differences of the energy isosurfaces when the energy centres of the hopfions approach to each other and the interference is bigger are a consequence of the change on the angular momentum. In the case of a total angular momentum equal to zero, the rotational direction of the Poynting vector of one hopfion is counterclockwise, while for the other hopfion is clockwise, along the $y$-axis, so the isosurfaces appear twisted. When the angular momentum is the same for each hopfion the rotational direction of the Poynting vector is the same for both and the isosurfaces appear without twist. 

We have provided some insight on the topology of the flow associated to the field lines (electric and magnetic ones) using the concept of electromagnetic helicity. By reversing the linear or the angular momentum of a hopfion, the electromagnetic helicity changes as the helicity of a photon would change. Therefore, one of the collisions considered in this work has non-null total electromagnetic helicity while the other case has zero electromagnetic helicity. For electromagnetic fields in vacuum, it is possible to define magnetic and electric helicities which are related to the mean value of the linking number of magnetic and electric lines respectively. Using the fact that the electromagnetic helicity is invariant and it is the sum of the magnetic and electric helicities, we have proved that in one case, the exchange mechanism \cite{Arr12} is acting between them, so they do not remain constant, while in the other case the helicities are constant.  

The results presented here may be of relevance for other solitons-like structures appearing in superfluid Helium \cite{Volovik1977}, Bose-Einstein condensates \cite{Kawaguchi2008} and ferromagnetic materials \cite{Dzyloshinskii}. 

\section*{Acknowledgements}
We want to thank Prof.~J.~M.~Montesinos Amilibia for useful discussions and guidance on Topology and Knot Theory matters during the writing of this work. This work was supported by research grants from the Spanish Ministry of Economy and Competitiveness (MINECO/FEDER) ESP2013-48032-C5-2 and (MINECO/FEDER) ESP2015-69909-C5-4.

\section*{References}
\bibliography{bibliobv1}

\end{document}